\pdfoutput=1

\documentclass[11pt]{article}

\usepackage[preprint]{acl}

\usepackage{times}
\usepackage{latexsym}

\usepackage[T1]{fontenc}

\usepackage[utf8]{inputenc}

\usepackage{microtype}

\usepackage{inconsolata}

\usepackage{graphicx}

\usepackage{algorithm}
\usepackage{algpseudocode}
\usepackage{amsmath}

\usepackage{parskip}
\title{Prompt Optimization and Evaluation for LLM Automated Red Teaming}

\author{
  \textbf{Michael Freenor}$^{1*\dagger}$, 
  \textbf{Lauren Alvarez}$^{1 2\dagger}$, 
  \textbf{Milton Leal}$^{1\dagger}$, 
  \textbf{Lily Smith}$^{1\dagger}$ \\
  \textbf{Joel Garrett}$^{1}$,
  \textbf{Yelyzaveta Husieva}$^{1}$, 
  \textbf{Madeline Woodruff}$^{1}$ \\
  \textbf{Ryan Miller}$^{1}$,
  \textbf{Erich Kummerfeld}$^{3}$, 
  \textbf{Rafael Medeiros}$^{4}$,
  \textbf{Sander Schulhoff}$^{5}$\\[1ex]
  $^{1}$Fuel iX Applied Research, Charlottesville, USA\\
  $^{2}$North Carolina State University, Raleigh, USA\\
  $^{3}$University of Minnesota, Minneapolis, USA\\
  $^{4}$TELUS Digital, Vancouver, CA\\
  $^{5}$Learn Prompting, San Francisco, USA
}

\begin{document}
\maketitle
\renewcommand{\thefootnote}{\fnsymbol{footnote}}
\footnotetext{$^*$Corresponding author: michael.freenor@fuelix.ai}
\footnotetext{$^\dagger$These authors contributed equally to this work.}

\begin{abstract}
Applications that use Large Language Models (LLMs) are becoming widespread, making the identification of system vulnerabilities increasingly important. Automated Red Teaming accelerates this effort by using an LLM to generate and execute attacks against target systems. Attack generators are evaluated using the Attack Success Rate (ASR) – the sample mean calculated over the judgment of success for each attack. In this paper, we introduce a method for optimizing attack generator prompts that applies ASR to individual attacks. By repeating each attack multiple times against a randomly seeded target, we measure an attack's discoverability – the expectation of the individual attack success. This approach reveals exploitable patterns that inform prompt optimization, ultimately enabling more robust evaluation and refinement of generators.
\end{abstract}

\section{Introduction}

\begin{figure*}[ht!]
    \centering
    \includegraphics[width=\textwidth]{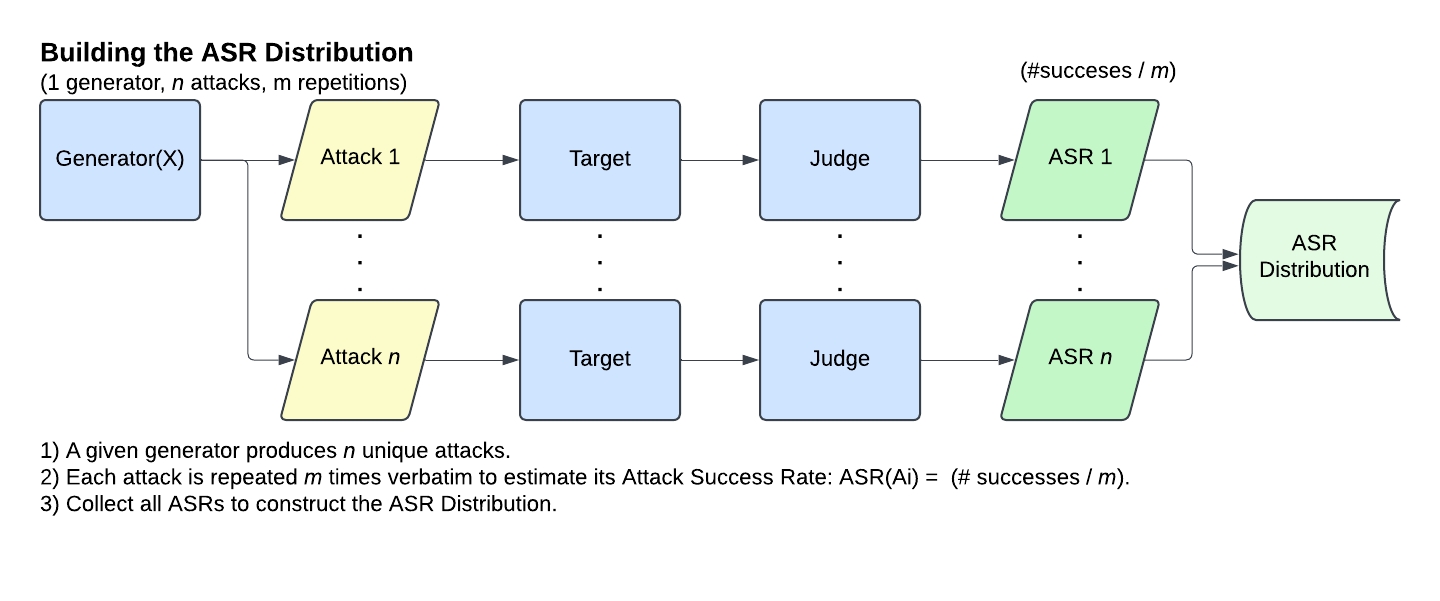}
    \vspace{-50pt} %
    \caption{Flowchart depicting the calculation of an empirical ASR distribution}
    \label{building_ASR_dist}
\end{figure*}

Systems that apply Large Language Models (LLMs), collectively referred to as \textit{LLM applications}, have demonstrated remarkable capabilities across various domains \citep{brown2020language}. However, like other forms of deep learning, these systems (\textit{targets}) are susceptible to adversarial inputs (\textit{attacks}). To address this, Red Teaming identifies critical vulnerabilities that malicious actors might exploit in real-world scenarios. This process reveals potential weaknesses and risks of LLM application misuse leading to liability for organizations encompassing social, moral, and legal dimensions \citep{yao2024survey}. Consequently, addressing these issues at scale through rigorous evaluation and optimization is paramount to ensuring scalability, alignment, safety, and security \citep{vashney2022trustworthy}.

Automated Red Teaming has emerged as a scalable approach to tackle this challenge \citep{perez2022attack}. In this practice, adversarial inputs are automatically generated by a prompted LLM (an attack generator) and systematically evaluated. Common examples of attacks include prompt injections and data exfiltration attempts \citep{dong2024attacks}. These attacks are deployed against a target and assigned a binary success feature (0 for failure, 1 for success), and the generator's Attack Success Rate (ASR) is calculated as the sample mean of these features \citep{Yang2023Entropy}.

One of our contributions (see Figure~\ref{building_ASR_dist}) is the application of ASR to individual attacks, replacing binary success with a numeric feature calculated over repeated trials of each attack against a randomly seeded target. This new set of features forms an \textit{ASR distribution} that captures more information about a given attack generator. Our main contribution is providing a method to optimize the attack generator through Optimization by PROmpting (OPRO) \citep{yang2024largelanguagemodelsoptimizers}. The in-context data for our OPRO procedure is formed by selecting semantically similar attacks that have a difference in ASR above a set threshold. These contrastive pairs exhibit subtle differences that can guide an LLM to produce more successful attack generators.

In the following sections we detail our methodology, present experimental results, and discuss the implications of our findings for the broader field of LLM safety and robustness.

\section{Related Work}

\subsection{Automated Red Teaming for LLM Applications}

Automated red teaming for Large Language Models (LLMs) has emerged as a critical area of research, with recent work establishing methods for systematically testing model safety and robustness. Early approaches focused primarily on gradient-based attack optimization \citep{bai2022constitutional}, but the field has rapidly evolved to encompass a broader range of techniques including automated discrete prompt-based methods \citep{perez2022redteaming, ganguli2022redteaming}, evolutionary algorithms \citep{chouldechova2023redteaming}, and hybrid approaches \citep{lin2024against}.

Recent taxonomies help organize these diverse approaches. \citet{perez2022redteaming} introduced a method for using one LLM to generate test cases for another demonstrating the potential of automated red teaming. \citet{ganguli2022redteaming} expanded on this work, providing insights into scaling behaviors and lessons learned from extensive red teaming efforts. These studies laid the groundwork for more sophisticated approaches.

In terms of implementation strategies, continuous optimization in embedding space has shown particular promise for generating nuanced adversarial inputs \citep{bai2022constitutional}. These gradient-based approaches operate in token embedding, hidden state, and output logit spaces, offering different trade-offs between attack power and computational complexity. However, discrete prompt-based methods remain dominant due to their universal applicability and interpretability \citep{perez2022redteaming, ganguli2022redteaming}. Furthermore, gradient-based exploits rely on access to powerful computational resources; for those using third-party services (e.g. OpenAI), these methods are off the table.

A significant advancement in automated red teaming has been the development of mutation-based approaches. These methods employ evolutionary algorithms where successful prompts are iteratively modified and selected based on their effectiveness \citep{chouldechova2023redteaming}. Advanced mutation techniques incorporate linguistic knowledge and adaptive mutation rates, while balancing effectiveness with attack diversity \citep{lin2024against}.

\subsection{Attack Success Rate}

The evaluation of red teaming effectiveness has coalesced around Attack Success Rate (ASR) as a primary metric, though its definition and application may vary across studies. ASR typically measures the proportion of attempts that successfully breach a model's safeguards \citep{hui2023language}, but recent work has highlighted the need for more nuanced interpretations.

\citet{dong2024attacks} propose a multi-dimensional view of ASR that considers not only the binary success outcome but also the severity of the violation and the naturalness of the attack. This approach has been further refined by incorporating semantic preservation scores to ensure that successful attacks maintain their intended meaning while achieving their objectives \citep{lin2024against}.

The importance of considering ASR in the context of attack diversity has been emphasized in recent literature. Some researchers have demonstrated that high ASR values may mask a lack of diversity in successful attacks, leading to potential overestimation of vulnerability \citep{chouldechova2023redteaming}. This has led to the introduction of metrics like the Attack Diversity Index (ADI) that should be considered alongside ASR when evaluating red teaming effectiveness.

As the field of LLM red teaming continues to evolve, there is a growing recognition of the need for standardized benchmarks and evaluation frameworks \citep{harmbench2024}.

\subsection{Optimization by Prompting (OPRO)}

Recent research has explored the potential of LLMs not just as generators of text or solvers of specific tasks, but as general-purpose optimizers. \citet{yang2024largelanguagemodelsoptimizers} introduced the concept of "Optimization by PROmpting" (OPRO), a novel approach that leverages the reasoning capabilities of LLMs to solve optimization problems across various domains.

OPRO represents a significant shift in how we utilize LLMs, moving beyond manual prompt engineering to employ these models as iterative optimizers. The core idea is to frame optimization problems as natural language tasks, allowing the LLM to propose, evaluate, and refine solutions through a series of prompted interactions.

The OPRO framework consists of several key components:
\begin{enumerate}
    \item Problem Formulation: The optimization task is expressed in natural language, including the objective function and any constraints.
    \item Solution Generation: The LLM proposes candidate solutions based on the problem description and previous iterations.
    \item Evaluation: The proposed solutions are evaluated using the specified objective function, often by an external system or LLM judges.
    \item Feedback and Iteration: The evaluation results are fed back to the LLM, which then reasons about how to improve the solution in the next iteration.
\end{enumerate}

Yang et al. demonstrated the effectiveness of OPRO on a diverse set of problems, including linear regression, traveling salesman problems, and prompt optimization for other AI tasks. Notably, OPRO showed competitive performance against traditional optimization algorithms and specialized neural network approaches, highlighting the versatility of LLMs as general-purpose problem solvers. 

One of the key advantages of OPRO is its ability to use the broad knowledge and reasoning capabilities embedded in LLMs. This allows the method to potentially discover novel optimization strategies that might not be apparent to human experts or easily encoded in traditional algorithms, especially for natural language inputs like adversarial prompts. For the purposes of this paper we employ OPRO to optimize attack generator prompts on the basis of an automated judge serving as the cost function.

\section{ASR Applied to Individual Attacks and its Distribution}

Recent work by \citet{chouldechova2023redteaming} has highlighted important theoretical considerations in AI Red Teaming, particularly regarding the interpretation and comparison of ASR across different contexts. Building upon this foundation, we propose an approach to address the specific challenges posed by non-deterministic LLM responses in Red Teaming scenarios. Our contribution complements existing work by offering a statistical method tailored to the unique characteristics of LLM-based systems.

\subsection{Applying ASR to Individual Attacks}

Traditional approaches to measuring attack success often rely on what we call \textit{single-try ASR}, which uses binary success on a single try of the attack as the feature for ASR calculations. This method provides a point estimate that serves as a basis of comparison between attack generators, benchmark datasets, and target systems \citep{chakraborty2021survey}. However, as pointed out by \citet{chouldechova2023redteaming}, this approach can be misleading in practice, particularly when dealing with non-deterministic LLM applications. Applying ASR to individual attacks provides additional information about the consistency and reliability of each attack across different target seeds, revealing important patterns that single-try measurements cannot capture.

We claim that \textit{attack quality} is at least partially characterized by this reapplied ASR calculation. Our primary justification follows from attacker economics: attackers operate under constraints of limited time and resources (or equivalently, limited tokens), making \textit{discoverability of success} an important part of overall quality \cite{chouldechova2023redteaming}. When attempting to uncover vulnerabilities, any repetition of previously executed attacks necessarily reduces the set of unique attacks tested within a fixed budget of trials. Consequently, attackers must balance between breadth and depth.

Targets typically operate with a random seed to maintain flexible and dynamic responses. Consequently, an identical attack executed multiple times may produce target outputs that vary, yielding some proportion of successes and failures that are not accounted for in single-try ASR. This observation aligns with the measurement theory concerns raised in previous work (Chouldechova et al.) and motivates the application of ASR to individual attacks.

We suggest the following pipeline:
\begin{enumerate}
    \item  Generate $n$ attacks from the attack generator.
    \item  For each unique attack, run the attack against the target (with no chat history) $m$ times.
    \item Collate the binary vector of m successes and failures produced by the judge with its corresponding attack and calculate the ASR (sample mean) of this vector.
\end{enumerate}

This approach replaces the typical binary quality metric with a numeric one, enabling a more informed assessment of generator quality based on the proportion of successes an attack achieved against a target \citep{Yang2023Entropy}. 

Given these considerations, our primary research goal is to demonstrate the utility of applying ASR to individual attacks and to show that its distribution provides valuable information about generator quality. We address this by exploiting the ASR distribution for generator prompt optimization.

\subsection{The ASR Distribution and Its Implications}

The ASR distribution provides critical information for evaluating and improving attack generators by revealing patterns that are invisible to single-try ASR. Specifically, this distribution:
\begin{itemize}
    \item Captures attack success discoverability across random target seeds
    \item Reveals clusters of particularly successful or unsuccessful attacks
    \item Enables identification of minimal linguistic differences between high and low-performing attacks
    \item Provides a statistical foundation for more robust generator optimization
\end{itemize}

To empirically compute the ASR distribution, we use GPT-4o (2024-05-13) for all three components of our pipeline: (1) the attack generator that creates adversarial prompts, (2) the target system that simulates an LLM-based application being attacked, and (3) the judge that evaluates whether each attack attempt succeeded. Each component operated with temperature=1 and random seeds to emulate realistic non-deterministic behavior. We generated $n=384$ unique attacks and evaluated each $m=50$ times against the target (see Appendix~\ref{sec:sample-sizes} for sample size justification).

\begin{figure}[t!]
    \centering
    \includegraphics[width=\columnwidth]{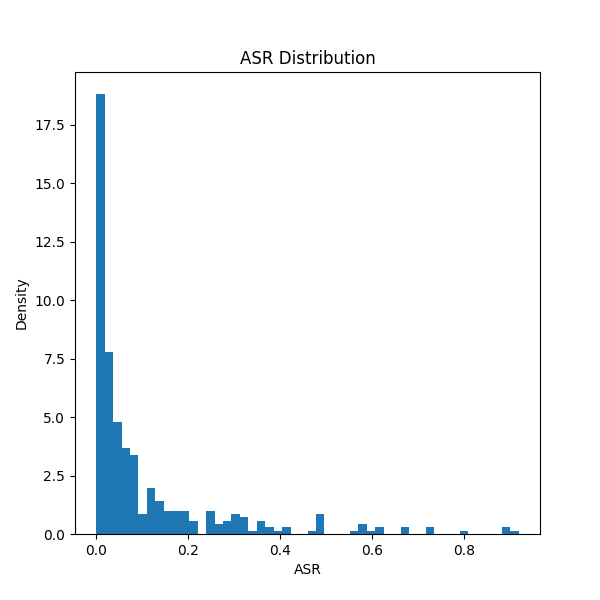}
    \caption{Example ASR Distribution}
    \label{fig:ASR_Dist}
\end{figure}

For the examined generator, we calculate an empirical distribution (Figure~\ref{fig:ASR_Dist}). This distribution is characterized by one mode near the origin, reflecting the effectiveness of the target's security measures, and additional mass higher on the ASR scale, representing a sample of relatively successful attacks.

To model this distribution, we suggest the use of a Beta mixture where the number of Beta distributions is equal to the number of modes present in the empirical distribution. This modeling approach serves several purposes: it provides a compact mathematical representation of the complex ASR patterns, enables statistical comparison between different attack generators, and facilitates simulation of attack success scenarios. Most importantly, it helps bridge empirical observations with theoretical understanding of attack quality, allowing researchers to make more informed predictions about generator performance in various contexts. 

The ASR distribution provides richer information for comparing generators than the traditional single-try ASR. While single-try ASR compares generators based on the probability that a randomly generated attack will succeed against the target on a single attempt, the ASR distribution is sensitive to settings where multiple tries are afforded to each attack. This distinction is crucial in practical red teaming scenarios as it integrates attack success discoverability.

Mathematically, the ASR distribution represents a distribution over Bernoulli parameters, with each attack's ASR serving as the $\theta$ parameter for its corresponding Bernoulli distribution. Thus, a set of single-try attempts across multiple attacks effectively samples from a mixture of these Bernoullis. Accordingly, the mean of the ASR distribution converges to the single-try ASR due to the fact that the ASR distribution ranges over a set of attack expectations. While two ASR distributions with equal means are equivalent on this basis, the one with more mass concentrated higher on the ASR scale would produce attacks whose success is more discoverable on the whole. This makes the ASR distribution a valuable tool for comparison whenever attack success discoverability is important.

In settings where the goal is exploiting a specific vulnerability, repeated trials of a set of attacks may be more desirable than trying a larger set of distinct attacks a single time each. The ASR distribution captures the nuance required for this assessment. Moreover, it highlights the noisy nature of binary success as a feature—low-ASR attacks may occasionally succeed on a single attempt while high-ASR attacks may fail, leading to potentially misleading classifications when using single-try metrics alone.

By leveraging this distributional understanding of generator quality, we can develop more effective strategies for both evaluating and improving attack generators. In the following section, we demonstrate how this information can be exploited to optimize generator prompts.

\section{Application: Attack Generator Optimization}
\subsection{ASR-Delta Pair Mining}

\begin{figure*}[ht!]
    \centering
    \includegraphics[width=\textwidth]{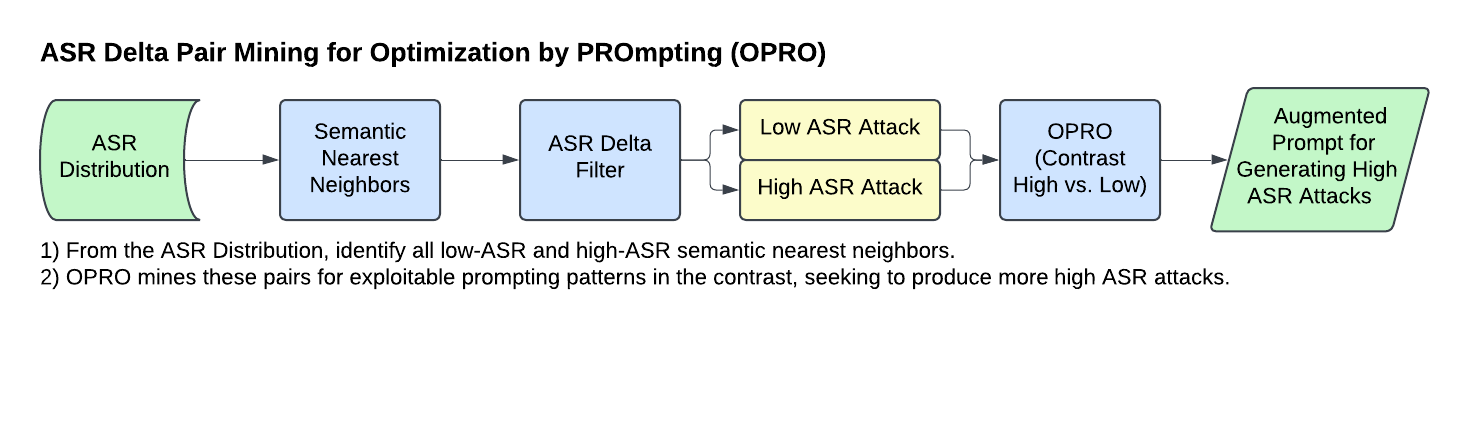}
    \vspace{-50pt} %
    \caption{Flowchart depicting ASR delta pair mining}
    \label{asr_dpm}
\end{figure*}

The ASR distribution reveals a diversity of attack quality that is not captured by single-try ASR. Without considering the ASR distribution, it is possible to overlook prompt patterns that correlate with the success of an attack. To demonstrate the practical utility of the ASR distribution, we explore an application to Automated Red Teaming using Optimization by PROmpting (OPRO), which is summarized in Figure~\ref{asr_dpm}. 

\begin{algorithm*}[t]
\caption{ASR-Delta Pair Mining and Generator OPRO}
\label{alg:CASD_OPRO}
\begin{algorithmic}[1]
\Require Set of distinct attacks $A$ generated by $G_A$ with ASR$(A) \sim D_A$, similarity metric $S$, threshold $\Delta > 0$, optimizer prompt $O$

\State Form set $P$ of semantically similar pairs $(a, b)$ where $a, b \in A$, $a \neq b$, and 
       \Statex \hspace{1.5em} $\forall a' \in A$ : $S(a, b) \geq S(a', b)$

\State Filter to pairs with significant ASR difference:
       \Statex \hspace{1.5em} $P' = \{(a, b) \in P \mid |$ASR$(a) -$ ASR$(b)| \geq \Delta\}$

\State Order each pair such that ASR$(a) >$ ASR$(b)$ for all $(a, b) \in P'$

\State Produce optimized generator $G_A' = O(G_A, P')$, where $O$ prompts for an addition 
       \Statex \hspace{1.5em} to $G_A$ that favors completions similar to high-ASR attacks $a$ rather than low-ASR
       \Statex \hspace{1.5em} attacks $b$
\end{algorithmic}
\end{algorithm*}

\begin{figure}[t!]
    \centering
    \begin{minipage}{0.495\textwidth}
        \centering
        \includegraphics[width=\textwidth]{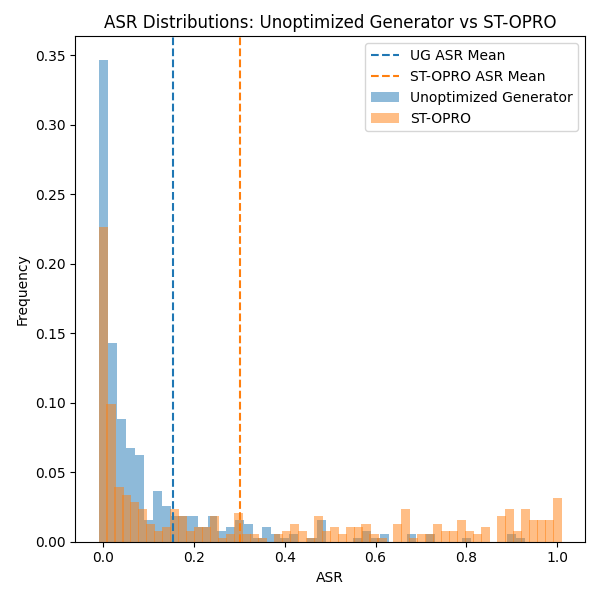}
        \caption{ASR Distributions: Unoptimized Generator vs ST-OPRO}
        \label{fig:OPRO_Comp1}
    \end{minipage}%
    \hfill
    \begin{minipage}{0.495\textwidth}
        \centering
        \includegraphics[width=\textwidth]{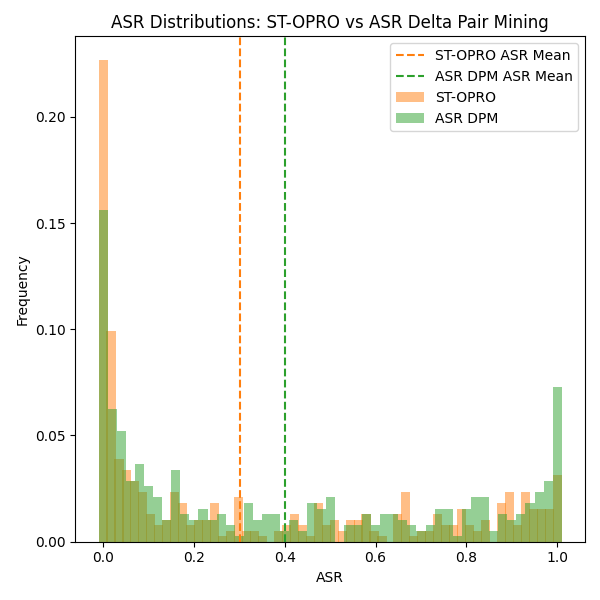}
        \caption{ASR Distributions: ST-OPRO vs ASR Delta Pair Mining}
        \label{fig:OPRO_Comp2}
    \end{minipage}
\end{figure}

This approach is formalized in Algorithm~\ref{alg:CASD_OPRO} which employs a technique we term \textit{ASR-delta pair mining}. This procedure selects similar attacks based on a chosen similarity measure (S). For S we apply cosine similarity to attack embeddings produced by OpenAI's text-embedding-3-large. We select attack pairs with an ASR difference exceeding a threshold $\Delta$. By setting an appropriate $\Delta$, this procedure can reveal critical and minimal contrasts between high and low-ASR attacks which prove effective for improving attack generator prompts against a given target. 

The core idea behind this approach is to provide semantic nearest neighbors with differing ASRs to an OPRO prompt, thereby highlighting which characteristics to emulate. The optimizer prompt is designed to produce text that, when \textit{added} to our generator's system prompt, transforms it into a more effective generator against the target.

In our experiments, we use GPT-4o (2024-05-13) with temperature=1 for both the generator and the optimizer. We produce 10 possible \textit{prompt additions} to the original generator and keep the one that yields the highest ASR distribution mean. This approach was chosen due to specific constraints in the prompt that are required to get attack generation to work, which can unintentionally be written out of a generator during optimization.

The result is demonstrated in Figure~\ref{fig:OPRO_Comp1} and Figure~\ref{fig:OPRO_Comp2}, which compares ASR distributions before and after the application of this method. The improvement in attack performance is evident in the rightward shift of density in the distribution, accompanied by a corresponding increase in the distribution's mean. This shift indicates that the optimized generator produces a higher proportion of high-ASR attacks using information from the original generator's ASR distribution. This approach:

\begin{itemize}
    \item Utilizes information about attack success discoverability from the ASR distribution
    \item Provides a systematic method for identifying and emulating successful attack characteristics
    \item Enables the continuous improvement of generators through iterative application
    \item Offers insights into the factors that contribute to attack success
\end{itemize}

However, it is crucial to consider potential limitations and areas for future research:

\begin{itemize}
    \item The generalizability of this method across different target systems needs further investigation
    \item The choice of similarity measure (S) and threshold ($\Delta$) may significantly impact results and thus require careful tuning
    \item The long-term effectiveness of this approach in the face of evolving defenses remains to be studied
\end{itemize}

By demonstrating the value of the ASR distribution as a source of information about attack quality, this work opens new avenues for research in automated red-teaming and LLM security. Future work could explore more sophisticated optimization techniques, investigate the transferability of learned improvements across different types of LLM-based systems, and develop methods to anticipate and counteract these optimized attack strategies.

\subsection{Method Comparison: Single-Try Attack Pair Mining}

To validate that the improvements observed in our optimized generator's ASR distribution stem from the information contained within the original generator's ASR distribution rather than from the act of OPRO itself, we conduct a comparative analysis. We compare ASR-delta pair mining OPRO against what we term \textit{single-try OPRO} (ST-OPRO). 

ST-OPRO replaces the use of individual attack ASR with the single-try binary success metric—selecting semantic nearest neighbors so long as one is successful and the other unsuccessful on a single try. This modification was specifically designed to provide the closest possible comparison to ASR-delta pair mining, differing only in the mechanism for distinguishing between high and low-quality attacks.

As in ASR-delta OPRO, we use GPT-4o (2024-05-13) with temperature=1 for both the generator and optimizer. We produce 10 possible prompt additions to the original generator and select the generator with the highest ASR distribution mean.

While ST-OPRO does yield improvements over the original unoptimized generator (see Figure~\ref{fig:OPRO_Comp1} and Figure~\ref{fig:OPRO_Comp2}), these improvements are demonstrably smaller than those achieved through the ASR-delta method.

The superior performance of ASR-delta pair mining suggests that the ASR distribution provides a better source of information for the optimization process than single-try success. This finding supports our hypothesis that the ASR distribution contains valuable information about attack quality that can be effectively leveraged for generator improvement.

\section{Conclusion}

This work demonstrates that single-try ASR, while useful as a point estimate for comparing attack generators, fails to capture potentially exploitable distributional information. We argue that binary success does not align as effectively with attack quality—defined through the lens of attacker economics—as our proposed application of ASR. This finding has implications for classification tasks, optimization procedures, and evaluation of generators.

Our results demonstrate that the ASR distribution serves as a reliable guide for identifying contrastive pairs for use with OPRO. By selecting semantic nearest neighbors with large ASR differences, we can isolate near-minimal contrasts in attack language that have disproportionate impacts on attack efficacy. This approach provides a method for understanding and improving attack generator performance, as evidenced by the superior results of ASR-delta pair mining compared to ST-OPRO.

These findings open several promising avenues for future research:

\begin{enumerate}
    \item Optimization of system prompts for more effective red teaming
    \item Generation of new insights about prompt engineering in security contexts
    \item Study of attack robustness across different LLM-based systems
    \item Development of more sophisticated defense mechanisms based on ASR distribution analysis
\end{enumerate}

Furthermore, this work suggests that the evaluation and optimization of LLM-based security systems benefit from more nuanced metrics of attack quality. The ASR distribution offers a more comprehensive view of attack success, enables more accurate predictions when attacks are repeated, and provides a method for evaluating and optimizing generators that is sensitive to attack success discoverability.

\section*{Limitations}

While our approach demonstrates promising results for optimizing attack generators, several limitations should be acknowledged:

1. Our experiments were conducted using a specific model (GPT-4o) and may not generalize to all LLM architectures.

2. The effectiveness of ASR-delta pair mining depends on the quality and diversity of the initial attack generator output.

3. The computational cost of repeatedly testing each attack multiple times may be prohibitive in some contexts, especially with limited API access or computational resources.

4. The proposed method focuses on optimizing for attack success rate but doesn't necessarily account for other important aspects such as attack stealth, naturalness, or transferability.

5. As defense mechanisms evolve, the effectiveness of optimized attack generators may diminish over time, necessitating continuous refinement.

\section*{Acknowledgments}

We thank our colleagues Christopher Frenchi and Nish Tahir for their helpful feedback throughout this process. Funding for this work was provided by Fuel iX, a TELUS Digital product.

\bibliography{main}

\clearpage

\appendix

\section{Computational Considerations for Sample Sizes}
\label{sec:sample-sizes}

In this study, we employed $n=384$ unique attacks, each evaluated $m=50$ times against the target. These sample sizes were selected based on both statistical considerations and computational feasibility.

\subsection{Determining the Number of Attacks ($n$)}

For the number of unique attacks $n$, we applied the standard sample size formula for estimating a population proportion with a specified margin of error:

\[n = \frac{z^2 \cdot p \cdot (1-p)}{e^2}\]

Where:
\begin{itemize}
    \item $z$ is the z-score corresponding to the desired confidence level (1.96 for 95\% confidence)
    \item $p$ is the estimated proportion (we used $p=0.5$ to maximize the required sample size)
    \item $e$ is the desired margin of error (0.05 or 5\%)
\end{itemize}

Substituting these values:

\[n = \frac{1.96^2 \cdot 0.5 \cdot 0.5}{0.05^2} = \frac{0.9604}{0.0025} = 384.16\]

This calculation yields $n \approx 384$, ensuring that our estimate of the overall ASR has a margin of error no greater than 5\% with 95\% confidence.

\subsection{Determining the Number of Repetitions ($m$)}

For the number of repetitions $m$ per attack, we selected $m=50$ primarily for computational efficiency. A full evaluation would require $n \times n = 384 \times 384 = 147,456$ total attack evaluations, which would be computationally prohibitive. Our approach with $m=50$ required only $384 \times 50 = 19,200$ evaluations, representing an 87\% reduction in computational load.

Importantly, we observed convergence in our ASR estimates at $m=50$. Additional repetitions beyond this point produced negligible changes in the estimated ASR values for individual attacks, suggesting that 50 repetitions adequately captured the success probability distribution while maintaining computational tractability.

\end{document}